\documentclass[11pt]{article}
\usepackage[a4paper,margin=2.4cm]{geometry}
\usepackage{amsmath,graphicx,booktabs,threeparttable,natbib,microtype}
\usepackage[skip=4pt]{caption}
\usepackage[colorlinks=true,linkcolor=black,citecolor=black,urlcolor=blue]{hyperref}
\hypersetup{pdftitle={Comment on "Feasibility of the Phobos 1 Hypothesis for Dark Comet 1998 KY26"}, pdfauthor={Stefano Ivan Stinga}}
\setcitestyle{authoryear,round}

\title{Comment on ``Feasibility of the Phobos~1 Hypothesis for Dark Comet 1998~KY$_{26}$''}
\author{Stefano Ivan Stinga\\ \small Independent researcher\\ \small \href{mailto:stefano@stinga.jp}{\texttt{stefano@stinga.jp}} \quad ORCID: \href{https://orcid.org/0009-0005-3860-3944}{0009-0005-3860-3944}}
\date{}

\begin{document}
\maketitle

\begin{abstract}
\citet{Hibberd2026a} proposed that the 11~m near-Earth object 1998~KY$_{26}$ is the Soviet spacecraft Phobos~1, and \citet{Hibberd2026b} fitted a Lambertian spacecraft model to the shapes of its light curves, with each segment's zero point free. I test that model, with its published geometry and brightness ratios, against the calibrated photometry of \citet{SantanaRos2025}. For an upper bound I scale the ratios until the brightest surface is a perfect Lambertian reflector, apply no shadowing, add all 33 spheres at the bus brightness and include all orders of reflection; the model must reach the observed rotation-averaged brightness on every night. At the authors' spin axis it falls short by at least a factor of 6.3 on the nights they fitted if the photometry is on the $V$ scale, and 4.8 in a conservative variant based on the observed $r'$ and $R$ bands. Even a generous estimate for mirror-like surfaces leaves a shortfall of about 2 there. At any spin axis the published ratios fall short by about 2 with Lambertian surfaces in the conservative variant, which also drops nights with possible panel glints; read as albedos, they fall short by about 2.4 in $r$ in an approximate energy budget for scattering roughly symmetric about the Sun. Between two nights with one instrument and filter, at 5$^\circ$ and 101$^\circ$ phase, the object fades 13.8 times; at the authors' axis their fitted model fades less than 1.4 times and the bound about 2.2. As published, the Phobos~1 model cannot reach the calibrated brightness of 1998~KY$_{26}$. Some models remain open: strongly mirror-like or idealized white Lambertian surfaces at some other spin axes and, in $r$, non-Lambertian white surfaces with white panel backs at any axis. None has yet been fitted to the light curves, and the released code can be adapted to test one. The test does not address the orbital evidence or other artificial origins.
\end{abstract}

\section{Introduction}
1998~KY$_{26}$ is an 11~m near-Earth object with a 5.35~min rotation period and a high geometric albedo \citep{SantanaRos2025}. Its nongravitational acceleration has been modeled with radiation forces \citep{Farnocchia2025}, a thermal-infrared nondetection limits its size \citep{Beniyama2025}, and its spectrum and albedo point to an enstatite-like surface \citep{Tatsumi2026}. It was to be visited by the Hayabusa2\# extended mission in 2031, but in 2026 November JAXA abandoned the rendezvous because of failing ion engines.\footnote{JAXA announced the change on 2026 November 17: Hayabusa2 is to fly by 2015~BM$_{510}$ in 2028 April instead (JAXA press-briefing material on the Hayabusa2 extended mission, in Japanese; \url{https://www.jaxa.jp/projects/files/youtube/hayabusa2-extended-mission/jaxa_doc01_20261117.pdf}).}

\citet{Hibberd2026a} proposed that the object is the Soviet probe Phobos~1, launched on 1988 July~7 and lost on 1988 September~2 \citep{NSSDCA}. They found that two burns, of 0.61~km~s$^{-1}$ on 1988 September~2 and 1.28~km~s$^{-1}$ on 1996 May~19, would bring the probe onto the orbit of 1998~KY$_{26}$ and to the object's position along that orbit. \citet{Hibberd2026b} then built an explicit photometric model of the spacecraft (their Appendix~A and Table~7). They fitted it to the light curves of \citet{SantanaRos2025}, obtaining a good match to the light-curve shapes with a spin axis at ecliptic $(\lambda,\beta)=(151^{+1}_{-2},+11^{+12}_{-3})^\circ$. They describe the identification as ``strongly plausible but presently unproven'' (their Sect.~6) and present the model as a first-order approximation with stated limitations (their Sect.~A.7). Their fit has a reduced $\chi^2$ of about 1.02, and the panel breadth and radiator inclination of the model, both favored by parameter scans against the light curves, are consistent with photographs of the spacecraft.

Because the model is explicit, its brightness can be checked. In their fit each of the 22 light-curve segments has its own additive zero point, and the analysis is described as ``not an attempt to fit absolute magnitudes'' \citep[][Sect.~2]{Hibberd2026b}. The absolute brightness of 1998~KY$_{26}$ has, however, been measured. \citet{SantanaRos2025} obtained calibrated photometry over phase angles of 5--101$^\circ$. They published it, with the Sun and Earth vectors of every point, as Source Data in the format of DAMIT \citep{Durech2010}. \citet{Hibberd2026b} fitted 954 points in 22 segments, which matches the file without the two light curves of 2024 November~7 (14 points): 330 points from 1998 and 624 from 2024. Here I ask whether their model can be as bright as 1998~KY$_{26}$.

\citet[][Sect.~2.4]{Hibberd2026a} made an initial brightness comparison. They took a diameter of 10~m and an albedo of 1, which treats the spacecraft as a filled disk of 78.5~m$^2$, and obtained $H=25.5$, close to the value of 25.6 they adopted for 1998~KY$_{26}$. The detailed model of \citet{Hibberd2026b} allows a sharper comparison. It presents a silhouette of at most 26--27~m$^2$ from any direction, depending on positions that their Table~7 does not give (Sect.~\ref{sec:energy}). The calibrated absolute magnitude of 1998~KY$_{26}$, $H=26.13\pm0.16$ \citep{SantanaRos2025}, corresponds to a product of geometric albedo and area of $pA\approx49$~m$^2$: near opposition the object is as bright as a face-on, perfectly white Lambertian disk of that area. Below I make this comparison night by night with the model itself.

\section{Data}
The Source Data contain 24 light curves: three from 1998 June and 21 from eight nights between 2024 May~19 and November~7. All are flagged as relative in the file (code 0 in the DAMIT format), but the intensities are reduced to unit distances from the Sun and the Earth, as the format specifies for calibrated curves. The mean magnitudes of the 2024 curves lie within $-0.16$/$+0.23$~mag of the published phase curve, $H=26.13\pm0.16$ with $G_1=-0.15$ and $G_2=1.0$ \citep{SantanaRos2025,Muinonen2010}. A fit to the eight 2024 nights alone, with equal weight per night, gives $H=26.128$. This shows only that the file is on their scale; it is not an independent check of their calibration. \citet{SantanaRos2025} describe their 2024 photometry as absolute, measured against non-variable solar-analog reference stars in each field, so I take the file to be on the published absolute scale. The 1998 curves, which they treated as relative, are not used in the brightness comparison.

For each 2024 night I pool all points with equal weight per point. A Fourier series in rotation phase (orders 1--4, period 5.3516~min) changes these means by about 1\% or less, except on November~7, whose 14 points give $-1$\% to $+4$\%. From the scatter about the second-order fits, the statistical errors of the means are 2\% or less, except on June~12 (3\%) and November~7 (5\%). I convert the mean reduced intensity $I$ (the file's brightness column, $10^{-0.4m}$ for reduced magnitude $m$) into
\begin{equation}
X_{\rm obs}=pA\,\Phi(\alpha)=\tfrac{\pi}{4}(1329~{\rm km})^2\,I,
\end{equation}
the product of geometric albedo, projected area and phase function, using the standard $V$-band relation \citep{PravecHarris2007}. Table~\ref{tab:nights} lists the values.

The file does not state the photometric band. According to the Methods of \citet{SantanaRos2025}, the 2024 light curves were taken in Sloan $r'$ (DECam, GTC/OSIRIS, Gemini/GMOS), except those from VLT/FORS2 (October~27 and November~7), which used R\_SPECIAL; additional $g'r'i'z'$ cycles were taken for colors. Their Supplementary Table~1 lists the Gemini night of May~31 as $griz$; I treat its light curve as $r'$. I consider three possibilities for the band.
\begin{itemize}
\item \textbf{$V$:} the file is on the $V$ scale used for the published albedo (nominal).
\item \textbf{$r$:} every night is Sloan $r'$. Converting to $V$ multiplies $X_{\rm obs}$ by 0.787. This uses $V-r=0.26$, as given by the transformation of \citet{Lupton2005} for $g-r=0.63$ \citep{Bolin2025}; that of \citet{Jester2005} gives 0.25.
\item \textbf{filter:} the two FORS2 nights, taken in R\_SPECIAL, are approximated as Cousins $R$, and the others are Sloan $r'$. The factors are 0.661 and 0.787, using $V-R_C=0.45$, as given by the transformations of \citet{Lupton2005} for the same $g-r$ and $r-i=0.15$ \citep{Bolin2025}; that of \citet{Jester2005} gives 0.38.
\end{itemize}
These transformations were derived from stars, so for an asteroid they are approximate. I adopt Lupton's, which give the larger $V-r$ and $V-R_C$ and so the lower $X_{\rm obs}$.\footnote{Before rounding, Lupton's $V-r$ is 0.2618, which would lower $X_{\rm obs}$ on the $r'$ nights by another 0.17\% and the values in Table~\ref{tab:results} by at most 0.01; no conclusion would change. His alternative relation for $R$, in $g-r$ alone, gives $V-R_C=0.47$ and would lower the FORS2 values by a further 2\%. This has no effect on the results: October~27, the only FORS2 night kept in the conservative variant (Sect.~3), sets none of the conservative values in Table~\ref{tab:results}.}

\begin{table}[htb]
\centering\small
\caption{Rotation-averaged brightness of 1998~KY$_{26}$ on the eight 2024 nights, from the Source Data of \citet{SantanaRos2025}, with $X_{\rm obs}$ on the $V$ scale (Sect.~2). The column $\sigma$ gives the statistical error of the nightly mean; errors in the nightly zero points and in the absolute calibration are not included (Sect.~\ref{sec:other}). The last column gives the Sun--Earth bisector (ecliptic J2000), the direction a flat surface must face for a specular glint to reach the Earth.}
\label{tab:nights}
\begin{threeparttable}
\begin{tabular}{lrrrrc}
\toprule
Night (UT) & Points & $\alpha$ ($^\circ$) & $X_{\rm obs}$ (m$^2$) & $\sigma$ (\%) & Bisector $(\lambda,\beta)$ ($^\circ$)\\
\midrule
2024-05-19 & 57 & 27.4 & 29.1\phantom{0} & 1.1 & (72, $-7$)\\
2024-05-21 & 177 & 30.0 & 26.9\phantom{0} & 0.6 & (75, $-7$)\\
2024-05-31 & 85 & 61.1 & 14.1\phantom{0} & 0.8 & (102, $-6$)\\
2024-06-05 & 161 & 82.3 & 6.84 & 1.1 & (117, $-4$)\\
2024-06-12 & 67 & 101.4 & 3.19 & 3.3 & (134, 0)\\
2024-10-27 & 37 & 10.7 & 44.9\phantom{0} & 1.9 & (221, $+3$)\\
2024-11-01 & 40 & 5.3 & 43.9\phantom{0} & 1.5 & (221, $+3$)\\
2024-11-07\tnote{$a$} & 14 & 5.3 & 47.7\phantom{0} & 5.0 & (222, $+3$)\\
\bottomrule
\end{tabular}
\begin{tablenotes}\footnotesize
\item[$a$] Not among the points fitted by \citet{Hibberd2026b}, as inferred from their point counts (Sect.~1) and supported by the exact reproduction of their panels-only $\chi^2$ (Sect.~\ref{sec:fidelity}).
\end{tablenotes}
\end{threeparttable}
\end{table}

\section{Model and test}
I use the Phobos~1 model of \citet[][Appendix~A, Table~7]{Hibberd2026b} without changes to its geometry, except for the alternative span below. Section~\ref{sec:fidelity} shows that the code, with shadowing added, reproduces their light-curve fit. The model has:
\begin{itemize}
\item two coplanar solar panels (inner and outer radii 2.40 and 5.00~m, width 2.50~m, 13.0~m$^2$) with white inner-edge stripes 0.60~m deep;
\item a cylindrical bus (radius 1.14~m, height 2.19~m);
\item two radiators ($1.60\times0.55$~m, inclined 0$^\circ$ and 35$^\circ$) and a high-gain antenna (disk of radius 0.60~m, canted 70$^\circ$ from the spin axis);
\item 33 spheres for tanks, torus and engine (17.8~m$^2$ of cross-section in total).
\end{itemize}
The body $z$ axis is both the spin axis and the panel normal, as in their Eqs.~9--12 (Sects.~A.2--A.3). All surfaces are Lambertian, as in their model, whose only specular term, for the upper spheres, was fitted to effectively zero ($\eta_{\rm metal}=1.25\times10^{-7}$, their Eq.~25). For Lambertian surfaces of reflectance $R_j$,
\begin{equation}
X_{\rm model}=\sum_j R_j\int\mu_0\,\mu\,dA,
\end{equation}
with $\mu_0$ and $\mu$ the cosines of the incidence and emission angles, and the sum averaged over rotation. A face-on Lambert disk then has $X=RA$ and a Lambert sphere $p=2R/3$. Their Table~7 does not give the height of the bus relative to the panel plane, the heights of the antenna and the radiators, or the clock angles of the lower tanks and of the torus spheres. In this unshadowed model these unspecified positions affect only the reflections between components and the silhouette. The adopted positions are documented with the code.

As an alternative to the 10~m span of their Table~7, I also use 12~m, the span listed in GCAT, where it is marked as uncertain \citep{McDowell}. With the inner radius and width of their Table~7 this gives 18~m$^2$ of panels.

For Lambertian surfaces whose reflectance cannot exceed 1, four choices make $X_{\rm model}$ an upper bound on the brightness of their model.
\begin{itemize}
\item The published ratios are scaled until the brightest surface reflects all the light that falls on it (case~A below).
\item No shadowing or occultation is applied.
\item All 33 spheres are added at the bus brightness, although \citet{Hibberd2026b} fold the light of all but the four upper spheres into the bus coefficient.
\item Reflections between components are included to all orders (Sect.~\ref{sec:multi}).
\end{itemize}
The test is $X_{\rm obs}\le X_{\rm model}$ on every night used. I report the shortfall $X_{\rm obs}/X_{\rm model}$ on the worst (binding) night. Where the spin axis is free, I minimize the shortfall over spin-axis directions, or poles (Sect.~\ref{sec:multi}). A value above 1 for every direction means that the model cannot reach the observed brightness at any spin axis. Results are given in two variants: nominal ($V$ band, all eight nights, no glint cut) and conservative (the ``filter'' band conversion of Sect.~2, 2024 November~7 dropped, and nights with a possible panel glint dropped). Some values are also given for the $r$ option of Sect.~2 (``in $r$'').

I set the reflectances in four ways.
\begin{description}
\item[A.] The published ratios relative to the panels: stripes 8 (their Eq.~21); radiators $\eta_{\rm board}q_{\rm bus}=4.218$, with $\eta_{\rm board}$ their radiator factor; bus barrel and top, antenna rear and spheres $q_{\rm bus}=3.1203$; antenna front $q_{\rm ant}=0.472$ (their Eq.~25). These are the ratios behind the published light-curve fit. \citet{Hibberd2026b} call them ``effective photometric coefficients rather than measured material albedos'' (their Sect.~2). Each multiplies the Lambertian flux of its surface relative to the panels (their Eq.~15), so for the areas of their Table~7 they act as the model's reflectance ratios, except that $q_{\rm bus}$ also carries the light of the tanks of the autonomous propulsion unit (ADU), the torus and the engine, which I add separately as spheres. As in their Sect.~A.3, the bottom of the bus is dark (they consider it largely obscured).\footnote{Giving the bus bottom the bus brightness would lower the values at the authors' axis, where it is never seen from the Earth and acts only through reflections, by about 0.1\% in case A and 1\% in case B, and the minima over spin axes by up to about 2\% and 5\%.} Scaling the ratios up until the brightest surface, the stripes, is a perfect white reflector ($R=1$) gives the brightest version of the model in which no reflectance exceeds 1; it puts the solar cells at $R=0.125$, within the range of real cells (see case C).
\item[B.] As A, but without the stripe enhancement, which was set by a separate scan and held fixed in the final fit. The radiators reach $R=1$.
\item[C.] The fitted ratios are abandoned in favor of idealized materials: solar cells $R=0.25$ on both faces, every other surface $R=1$. Space solar cells absorb 75--91\% of sunlight, and common white spacecraft paints (S-13G, Z-93) reach a solar reflectance of $R\simeq0.80$--0.83 when new \citep{Henninger1984}; such coatings tend to darken during years in space.
\item[C2.] As C, but with cells on one face only and the other face of the panels a perfect white reflector. Both choices of the cell face are tested.
\end{description}

For twelve other sets of the unspecified positions (``placements'') among those of Sect.~\ref{sec:energy}, the values reported below change by less than 1\% for case A and by about 2\% for case B. For idealized materials (cases C and C2), where reflections matter more, the values are up to about 8\% lower at the authors' axis and up to about 23\% lower at other spin axes (Sect.~\ref{sec:results}). Unless stated otherwise, values in the text and in Table~\ref{tab:results} are for the adopted positions. Shortfalls are rounded down and upper limits on power and area up. Unless the span is named, a single shortfall is the smaller of the 10~m and 12~m values.

\paragraph{Mirror-like reflection.} The model is Lambertian, but real solar panels carry cover glass and radiators may carry mirror coatings. Because the panel normal is the spin axis, a panel or the flat radiator can glint toward the Earth on a given night only when the spin axis lies within the specular lobe around that night's Sun--Earth bisector (Table~\ref{tab:nights}). The three low-phase nights share almost the same bisector, $(\lambda,\beta)\approx(221^\circ,+3^\circ)$, so a spin axis near it or its opposite could add a glint on all three. Such a glint would be steady through each rotation and would dilute the light-curve amplitude. The observed amplitudes argue against it, but I do not rely on this argument.\footnote{To close the nominal shortfall of about 4 at such axes (Sect.~\ref{sec:results}), the glint would have to supply three quarters of the light, and the rest would then have to vary by more than twice its own mean to give the observed peak-to-trough range of 0.5--0.6 times the mean on 2024 October~27 and November~1. Noise inflates these ranges, and \citet{SantanaRos2025} used the fall data for their phase curve but judged them too faint to build light curves.} Instead I drop, for each spin axis, every night whose bisector lies within $15^\circ$ of the axis in either direction (one of the three steps of the conservative variant). I chose the width of the cut as a conservative allowance for unmodeled glints; it is not a measured lobe width. It allows for lobes several degrees wide and is generous for cover glass, whose specular reflectance is a few percent. Section~\ref{sec:other} gives the effect of other cuts.

The bus barrel, the curved side of the cylindrical bus, could glint in the same steady way when a night's bisector is nearly perpendicular to the spin axis (its glint condition). All eight bisectors lie within 7$^\circ$ of the ecliptic, so for spin axes near the ecliptic poles a barrel glint is possible on every night. With the published ratios the barrel reflects 39\% diffusely, which leaves at most 61\% for a specular part. A lobe with a sharp core and broad wings could then supply the missing light on every night at such axes (Sect.~\ref{sec:results}), but on the low-phase nights and on May~19 and 21 the glint would have to provide about three quarters or more of the light as a steady signal. The same dilution argument applies, but I set it aside as well.\footnote{On November~1, October~27 and May~19 the rest of the light would have to vary by two to three times its own mean in the conservative variant, and more in $V$, but individual points on these nights scatter by 8--12\%, which inflates the observed ranges.}

The spheres could also be partly mirror-like. A mirror sphere reflects light equally in all directions, so 61\% specular spheres would add a steady 2.7~m$^2$ on every night.

\section{Results}
\label{sec:results}
Table~\ref{tab:results} gives the bounds for both variants.

\begin{table}[htb]
\centering\small
\caption{Worst-night shortfall $X_{\rm obs}/X_{\rm model}$, including reflections between components. Nominal: $V$ band, all eight nights, no glint cut. Conservative: ``filter'' band conversion, with 2024 November~7 and any night with a possible panel glint dropped (Sect.~3). Cases (Sect.~3): A, published ratios; B, without the stripe enhancement; C, idealized materials; C2, as C with white panel backs. Surfaces are Lambertian except in the last three rows. Values for ``any'' spin axis are minimized over spin-axis directions. Values are rounded down and use the adopted positions where their Table~7 gives none (Sect.~3).}
\label{tab:results}
\begin{threeparttable}
\begin{tabular}{llcccc}
\toprule
 & & \multicolumn{2}{c}{10~m span (published)} & \multicolumn{2}{c}{12~m span}\\
\cmidrule(lr){3-4}\cmidrule(lr){5-6}
Case & Spin axis & nominal & conservative & nominal & conservative\\
\midrule
A\tnote{$a$} & authors' $(151^\circ,+11^\circ)$ & 6.81 & 4.91 & 6.74 & 4.86\\
B & authors' & 3.54 & 2.56 & 3.50 & 2.53\\
C & authors' & 2.33 & 1.68 & 2.32 & 1.67\\
C2 & authors' & 2.23 & 1.60 & 2.17 & 1.56\\
\midrule
A & any & 4.31 & 2.32 & 4.08 & 2.19\\
B & any & 2.72 & 1.45 & 2.55 & 1.36\\
C & any & 1.69 & 0.89 & 1.61 & 0.86\\
C2 & any & 1.30 & 0.70 & 1.16 & 0.61\\
\midrule
A, mirror-like spheres\tnote{$b$} & authors' & 4.90 & 3.54 & 4.87 & 3.51\\
A, mirror-like surfaces\tnote{$c$} & authors' & 3.8\phantom{0} & 2.7\phantom{0} & 3.6\phantom{0} & 2.6\phantom{0}\\
A, mirror-like spheres\tnote{$b$} & any & 3.46 & 1.82 & 3.31 & 1.74\\
\bottomrule
\end{tabular}
\begin{tablenotes}\footnotesize
\item[$a$] Without 2024 November~7, which \citet{Hibberd2026b} did not fit, the nominal values are 6.45 and 6.39, set by October~27.
\item[$b$] Spheres 61\% specular (Sect.~3).
\item[$c$] Spheres, bus barrel and flat surfaces (panels, bus ends and flat radiator) at their largest specular part, with single-bounce lobes that are symmetric about the mirror direction, decrease away from it and are chosen for each night. Values are approximate, so only one decimal is given. More generous assumptions lower the conservative value to about 2.0--2.1, possibly below 2 (see text).
\end{tablenotes}
\end{threeparttable}
\end{table}

\begin{figure}[t]
\centering\includegraphics[width=0.92\textwidth]{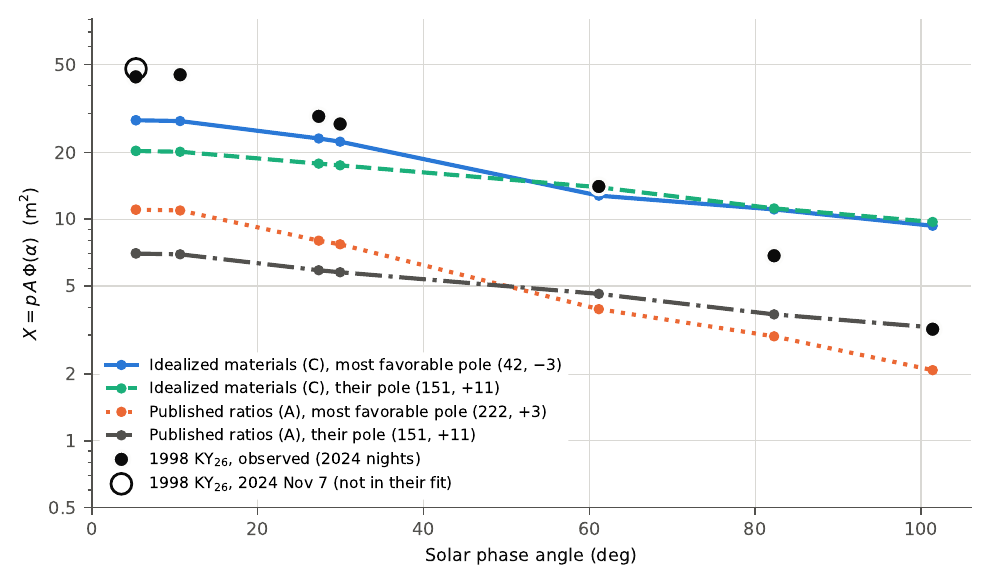}
\caption{Observed brightness of 1998~KY$_{26}$ on the eight 2024 nights (black, $V$ scale; open circle: November~7, which the authors did not fit) and upper bounds from the Phobos~1 model of \citet{Hibberd2026b}, including reflections between components (10~m span, Lambertian). Curves connect the model values at the same eight geometries. If the photometry is on the $V$ scale, a model consistent with the data must lie at or above every black point. Statistical errors (Table~\ref{tab:nights}) are smaller than the symbols. The ``most favorable'' poles in the legend minimize the shortfall in the nominal variant; both lie in the glint cones within which the conservative variant drops the low-phase nights.}
\label{fig:brightness}
\end{figure}

\begin{figure}[t]
\centering\includegraphics[width=\textwidth]{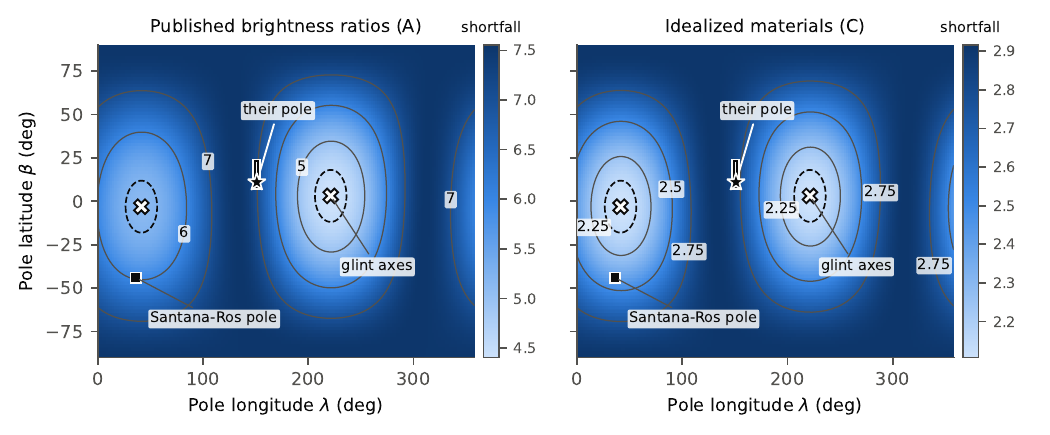}
\caption{Worst-night shortfall for every spin-axis direction (ecliptic J2000; direct light only, 10~m span, $V$, all nights, Lambertian). Reflections lower these values by 2--3\% in case A and 14--20\% in case C (adopted positions). The star marks the authors' pole; the narrow box around it is the 68\% region of that pole, $\lambda=149$--$152^\circ$ and $\beta=+8^\circ$ to $+23^\circ$. The square marks the pole of \citet{SantanaRos2025}, $(36^\circ,-44^\circ)$. The crosses mark the Sun--Earth bisector of the low-phase nights and its opposite, where the Lambertian minimum lies and where a specular glint from the panels could add light. The dashed circles are the 15$^\circ$ cones within which the conservative variant drops these nights. The two panels have separate color scales.}
\label{fig:polemap}
\end{figure}

\paragraph{The published solution.} At the authors' spin axis the model with the published ratios is at least 6.3 times too faint in $V$ on the nights they fitted (6.7 with November~7), and 4.8 times in the conservative variant (Table~\ref{tab:results}, note~$a$, and Fig.~\ref{fig:brightness}). In $V$ the shortfall is largest at low phase, about 6.1--6.8 at 5--11$^\circ$, and falls to about 1 at 101$^\circ$, where the bound just reaches the observed brightness; their model as fitted, with shadowing and without the added spheres, is 2.0 times too faint there when scaled in the same way (Sect.~\ref{sec:fidelity}). Across the 68\% region of their pole (Fig.~\ref{fig:polemap}), the model is at least 6.68 times too faint in $V$ with all eight nights, November~7 included, and 4.81 in the conservative variant. Equivalently, its stripes would need a Lambertian reflectance of nearly 5 to 7, against a physical limit of 1. Without the stripe enhancement (B) the model is at least 3.3 times too faint in $V$ on the nights they fitted (3.5 with November~7) and 2.5 in the conservative variant. Even the idealized materials of cases C and C2 fall short at this axis, by 1.67 and 1.56 in the conservative variant for the adopted positions, and by about 8\% less for the other placement most favorable to the model.

Without the three fall nights, which \citeauthor{SantanaRos2025}\ judged too faint for light curves (Sect.~3) and where a real surface could be brighter near opposition than a Lambertian one, the shortfall at this axis is still at least 4.9 in $V$ and 3.8 in the conservative variant, set by 2024 May~19 at 27$^\circ$ phase. On May~19 and 21, however, the barrel lies only 9--12$^\circ$ from its glint condition at this axis. If these two nights are dropped as well, the published ratios fall short by at least 2.9 in $V$ and 2.3 in the conservative variant, but the bound no longer excludes idealized materials.

At this axis the panel normal lies 70$^\circ$ from the bisector of every low-phase night, and at least 19$^\circ$ from those of the May--June nights. No night's bisector lies within 15$^\circ$ of the normal of the panels or of the flat radiator, so no night is dropped for glints. The inclined radiator and the bus barrel could glint only on May--June nights, which do not bind here when the fall nights are included. Surfaces whose normals turn with the rotation could flash. The high-gain antenna, canted 70$^\circ$ from the spin axis, sweeps through the low-phase bisectors once per rotation. To close the gap on November~1, a flash would have to supply about 80\% of that night's light within a few seconds of each 5.35~min rotation. A broader lobe would give a longer flash but deliver proportionally less light per rotation; only a nearly flat mirror could supply this much. Each 40~s exposure \citep[][Supplementary Table~1]{SantanaRos2025} covers an eighth of a rotation, so a frame that caught the flash would be six to seven times the light-curve mean and the others about a fifth of it. Over the 10--12 rotations of each night the exposures cover nearly every rotation phase, so several frames would have caught it. No frame on October~27 or November~1 exceeds 1.24 times the mean of its light curve or falls below 0.66 of it, so the photometry shows no such flash. Had such frames been rejected as outliers, the remaining frames would measure only the steady light, which the model would then have to reach on its own.

Mirror-like reflection does not close the gap either. With 61\% specular spheres the shortfall is at least 4.6 in $V$ on the nights they fitted (4.8 with November~7) and 3.5 in the conservative variant. No other single surface can glint steadily toward the Earth at this axis on the low-phase nights: the barrel lies about 20$^\circ$ from its glint condition and the flat surfaces (panels, bus ends and flat radiator) 70$^\circ$. Broad lobe wings could still add some light. If these surfaces and the spheres are given their largest specular part, with single-bounce lobes that are symmetric about the mirror direction, decrease away from it and are chosen separately for each night, the shortfall is about 3.6 in $V$ (with November~7) and 2.6 in the conservative variant. With more generous assumptions (lobes confined above the surface, the inclined radiator and the antenna rear included, and an antenna flash as large as the frames allow), the conservative value falls to about 2.3 with cells like cover glass, and to about 2.0--2.1 with cells that are 87.5\% specular (the most that a diffuse reflectance of 0.125 allows), far more mirror-like than cover glass. Asymmetric lobes, or a specular double bounce between the panels and the bus (2.6--8.7$^\circ$ from its glint condition on the low-phase nights), could lower it below 2, largely because of the 87.5\% specular cells.

\paragraph{Any spin axis.} \citet{Hibberd2026b} refitted $q_{\rm bus}$, $q_{\rm ant}$, $\eta_{\rm board}$ and $\eta_{\rm metal}$ at each trial pole (their Sect.~A.5), so a fit at another axis would give other ratios. The published ratios show what moving the axis alone does. Cases C and C2, with every surface other than the solar cells at $R=1$, cover other ratios for Lambertian surfaces with solar cells no brighter than $R=0.25$. With the published ratios and Lambertian surfaces the lowest shortfall in $V$, 4.0--4.3, occurs with the spin axis about 1$^\circ$ from the low-phase bisector (Fig.~\ref{fig:polemap}). Once the nights on which a panel glint could add light are dropped, the shortfall is at least 2.7 in $V$ for any axis, and 2.1 in the conservative variant (about 2 allowing for errors in the nightly zero points; Sect.~\ref{sec:other}). The lowest values lie just inside the glint cone of the low-phase nights, near $(\lambda,\beta)=(56^\circ,-6^\circ)$. There the binding night is 2024 May~19 (phase angle 27$^\circ$), whose bisector lies 16$^\circ$ from the axis, just outside the cut. Without the stripe enhancement (B) the conservative shortfall is at least 1.36. These values assume Lambertian surfaces, as in the model, and panels whose specular reflectance is that of cover glass.

Mirror-like spheres (61\% specular) lower the conservative minimum to 1.82 for the 10~m span (1.74 for 12~m). With a mirror-like bus, spin axes near the ecliptic poles are not excluded: for a suitable lobe the glint could supply the missing light on every night used, for example at $(285^\circ,-83^\circ)$. At spin axes near the low-phase glint direction, mirror-like spheres together with panels that have an 87.5\% specular part (far more than solar cells reflect) and a lobe broad enough to add light on the May--June nights also bring the shortfall to about 1.

\paragraph{Brightness of the spheres.} At the authors' axis the 33 spheres supply about two thirds of the case-A bound on the low-phase nights. In the fit the light of all but the four upper spheres is not a separate term but is absorbed into $q_{\rm bus}$ \citep[][Sect.~A.3]{Hibberd2026b}, so adding them at the bus brightness counts that light twice. At that axis the light curves allow no sphere light beyond what $q_{\rm bus}$ already carries: added to the published model as shadowed Lambertian reflectors, the spheres raise $\chi^2$ from 928.1 to 1152 at the bus brightness and to 1284 if they are white, and refitting $q_{\rm bus}$, $q_{\rm ant}$ and $\eta_{\rm board}$ together with them gives no sphere light at all ($\chi^2=923$, with the shadowed model of Sect.~\ref{sec:fidelity}). The photograph in their Fig.~2 shows the ADU tanks as white globes, but white spheres would make a different model: at this axis a refit of the other coefficients still gives $\chi^2=1077$, and at other axes it has not been fitted. With white Lambertian spheres the shortfall at the authors' axis would still be about 2.8 in $V$ on the nights they fitted and 2.1 in the conservative variant. At the best other spin axis it would be only about 1.2 in that variant. With white spheres, the energy bound of Sect.~\ref{sec:energy} would leave a shortfall of about 1.3 in $r$, allowing for its 9\% error.

\paragraph{Idealized materials.} With idealized materials instead of the fitted ratios (C), the lowest nominal shortfall over spin axes is 1.6 for the adopted positions. Dropping the glint nights leaves a shortfall of 1.09 in $V$. In the conservative variant it falls to 0.86--0.89, and the bound no longer excludes case C. With white panel backs as well (C2), the bound no longer excludes the model in $V$ once the glint nights are dropped (0.78--0.89). For the 12~m span the same holds in $r$ without dropping them (0.91). With idealized materials the reflections between components add 12--25\% at low phase, and they include light exchanged between the underside of the bus and the torus and tanks beneath it, which the unshadowed bound counts as sunlit and visible although they hide one another. These reflections depend on the unspecified positions: for the other placements tried, the values at the same spin axes are up to about 23\% lower for C and 19\% lower for C2, most when the underside of the bus just clears the torus. For the placement most favorable to the model, the nominal shortfall of case C falls to 1.25 and its value in $V$ without the glint nights falls below 1, so the shortfall of 1.09 given above is not significant. For C2 the nominal values of Table~\ref{tab:results} (1.30 and 1.16) fall to 1.06 and 0.96. For C and C2 the adopted positions give the highest values of the thirteen placements tried. These values for idealized materials are generous to the model, because the unshadowed bound counts surfaces hidden from the Earth. At the best C2 axis, for the 10~m span, the observed brightness on May~19 in $V$ (29.1~m$^2$) exceeds the whole area the spacecraft presents to the Earth (24.9~m$^2$ for the adopted positions and at most 26.1~m$^2$ for the other placements of Sect.~\ref{sec:energy}), so direct light alone could not reach it; for 12~m that area is 28--31~m$^2$, close to the observed value.

For both C and C2 the best spin axes lie at the edges of the glint cones, near $(236^\circ,+6^\circ)$ and $(56^\circ,-6^\circ)$, or close to the low-phase bisectors. All of them lie 69--84$^\circ$ from the published axis or its opposite, whichever is closer. For the adopted positions and Lambertian surfaces, the bound excludes idealized materials everywhere within 55$^\circ$ of either with the 15$^\circ$ glint cut, and within 56$^\circ$ without it. Near these limits the shortfall is only 1.0--1.3, and I have not repeated this search for other placements. If the bus is instead allowed to be a mirror and every night on which it could glint is dropped, the bound does not exclude an idealized model as close as 12$^\circ$ to the published axis, just outside its 68\% region. Only the three May--June nights at 61--101$^\circ$ phase then remain. On the other five the model without a glint is about 1.5--2.4 times too faint in $V$, so a glint from the bus would have to supply about a third to a half of their light or more, since a barrel that glints cannot also be a perfect diffuse reflector. Such a model would be new, and its light curves remain to be fitted.

\paragraph{Fading with phase angle.} The bound changes less with phase angle than 1998~KY$_{26}$ does (Fig.~\ref{fig:brightness}). Between 5$^\circ$ and 101$^\circ$ the observed brightness falls by a factor of 13.8. For direct light without shadowing, the bound falls by at most 8.5 for case A at any spin axis (10~m span; 8.9 for 12~m), at most 10.1 for idealized materials, and 2.2--2.3 at the authors' pole. Both nights, 2024 November~1 and June~12, were observed with GTC/OSIRIS in $r'$ \citep[][Supplementary Table~1]{SantanaRos2025}, so this comparison depends neither on the band conversion nor on the calibration of one telescope against another. The added spheres, whose light falls 4.4 times between these two nights, limit how much the bound can fade at spin axes where the rest of the model fades steeply. At the authors' axis they make it fade more, not less: their model as fitted, without the added spheres, fades only 1.1--1.2 times between these two nights without shadowing, and 1.33 times with it (10~m span), against 13.8 for the object. At other spin axes the model as fitted can fade more than the object does: without shadowing it does so at 12--15\% of all spin axes, none of them within 20$^\circ$ of theirs. The ratio therefore tests the model at the authors' axis but not at every axis. Away from opposition the ratio disagrees in the same way at the authors' axis: between May~21 and June~12 (30$^\circ$ and 101$^\circ$ phase), both also observed with GTC/OSIRIS in $r'$, the object fades 8.4 times, while the bound fades 1.7--1.9 times and their model as fitted, with shadowing, brightens slightly (10~m span).

This difference in fading is also why I compare the model with each night separately rather than fitting one overall scale. If one overall scale is fitted to all nights, the bound is too faint at low phase and too bright at high phase. At the authors' pole (10~m span, direct light, equal weight per 2024 night), the residuals after the best scale are about 0.7~mag rms, with 1998~KY$_{26}$ brighter by 0.55--0.64~mag at 5--11$^\circ$ and fainter by 1.4~mag at 101$^\circ$. A fit with one overall scale averages errors of opposite sign, so its result depends on how the nights are weighted.

\section{Robustness}
\subsection{Reproducing the published fit}
\label{sec:fidelity}
To check that the code implements the model of \citet{Hibberd2026b}, I compared it with a separate implementation of their flux (their Eq.~15), written from the text of their Appendix~A. With shadowing off and without the added spheres, the direct light used here equals that flux divided by 8, the case-A scaling, to within one part in $10^{12}$ at each of 1.4 million test geometries. I then added mutual shadowing and observer occultation by ray tracing, with the unspecified positions set as in this code. At their spin axis, with their coefficients, 22 segment zero points and one rotation phase per observing group, this model fits their 954 points with $\chi^2=928.1$ and an rms of 0.138~mag; they report 934.2 and 0.139~mag. The small difference lies within the range, $\chi^2=918$--944, spanned by the unspecified positions (948 if the light of the bus is itself shadowed and occulted by the other parts, rather than integrated without masking). Their panels-only control fit, $\chi^2=1506.42$, is reproduced exactly, which confirms the data selection and weights. Their panels-plus-bus control fit, $\chi^2=1231.65$, gives 1232.0 here. The fit also checks my reading of their Eq.~15, which the comparison of the two codes cannot: scaling the bus, the antenna and the radiators together by a factor of 2 relative to the panels raises $\chi^2$ from 928.1 to 968, so a reading that made these parts twice as bright is excluded, and a dimmer one would only lower the bound. Refitting their coefficients in this code lowers $q_{\rm bus}$ to 2.36 and $q_{\rm ant}$ to 0.27 ($\chi^2=923.0$), so the published values used here are, if anything, on the bright side. Shadowing shapes the light curves but removes only 1--17\% of the rotation-averaged light at their spin axis. Even without its reflections, the upper bound used here is 1.8--3.8 times brighter than the shadowed model on every 2024 night they fitted, mostly because of the added spheres. For comparison, their model as fitted (with shadowing and without the added spheres), scaled so that the stripes reflect all the light, is at least 21 times too faint in $V$ and 16 times in the conservative variant on the nights they fitted, so the bound used here is generous by a factor of more than three.

\subsection{Reflections between components}
\label{sec:multi}
Light reflected by one part of the spacecraft can reach another and then the observer. I compute this with a Monte Carlo ray trace of the same geometry.
\begin{itemize}
\item Every surface facing the Sun acts as a source, with no shadowing, at the actual Sun direction for each night and rotation phase.
\item All orders are summed until the sum converges, with a geometric estimate of the remainder.
\item Sample points buried inside overlapping solids of the model are excluded.
\end{itemize}
Reflections add about 3\% of the direct light in case A and about 8\% in case B on the low-phase nights and on May~19 and 21. At 101$^\circ$ they reach up to about 30\% in case A, depending on the spin axis. In cases C and C2 they add 12--25\% at low phase for the adopted positions.

Some minima lie in regions narrower than 1$^\circ$ between the edges of the glint cones, so the spin-axis search combines exact direct light with interpolated reflections:
\begin{itemize}
\item Direct light is evaluated on about 1.05 million directed axes: a global 0.25$^\circ$ grid, plus points 0.003$^\circ$ on either side of every cone boundary, 0.05$^\circ$ apart.
\item Reflections are computed on a global 12$^\circ$ grid and interpolated.
\item The ten best distinct candidates of each variant are refined on a 0.02$^\circ$ grid and recomputed with exact reflections.
\end{itemize}
At the candidates, interpolated and exact values differ by at most 0.1\%. The results agree with a higher-resolution calculation to within about 0.5\%. Both calculations credit each ray with the light of the nearest sample point on the surface it strikes. Where the underside of the bus is sunlit, this undercounts the light trapped between it, the torus and the tanks, so I also follow every ray from its actual hit point (a path-traced calculation), at the authors' axis and at every candidate. For idealized materials this lowers the shortfalls at the minima by up to about 0.5\%, and by up to about 2.5\% for placements in which the underside of the bus just clears the torus; for cases A and B the change is about 0.1\% or less. Table~\ref{tab:results} gives the lowest of the three calculations. The path-traced values have Monte Carlo errors of about 0.1\% (up to about 0.3\% for those placements); because the lowest value over the candidate axes is taken, the minima tend to be low, in the model's favor, by about 0.1\%. For cases C and C2, the values for other placements, the limit of 56$^\circ$ and the ranges of reflected light are path-traced.

\subsection{Energy}
\label{sec:energy}
A second, approximate bound needs no specific scattering law, only the assumption that the scattered light is roughly symmetric about the Sun direction. It therefore covers multiple reflections and scattering of that kind, such as strong backscattering near opposition, but not specular or forward-scattering lobes; glints are treated above. Under that assumption $pAq$, where $q$ is the phase integral truncated at 101$^\circ$, is a lower bound on the reflected power, in units of the solar flux. Integrating the phase curve fitted by \citet{SantanaRos2025}, extrapolated below 5$^\circ$, from 0 to 101$^\circ$ gives $pAq\ge35.4$~m$^2$ (35.4--35.8~m$^2$ for the three phase curves they considered). A log-linear interpolation through the eight nightly values, held constant below 5$^\circ$, gives at least 34.2~m$^2$, and I use this lower value, because the test of the assumption described next uses the same interpolation. The assumption is not exact for this geometry. When I apply the same estimate to the unshadowed Lambertian model itself, at the geometries of the eight nights and so including their changes of aspect, it exceeds the model's true maximum power by up to 9\% at some spin axes in case A, and 4\% in case C. The power that can leave the spacecraft is at most the light first striking surfaces other than the solar cells, plus at most 25\% of the light first striking the cells. Because the parts shadow one another, this maximum depends on the unspecified positions. I therefore take the largest value over all Sun directions and over 80 placements: five heights of the bus relative to the panel plane, two of the antenna, four of the radiators and two clock angles of the lower tanks, with the torus at clock 0 (all are listed with the code and scanned at low resolution, and the adopted placement and the extreme ones are recomputed at full resolution). It is 20.71~m$^2$ for the 10~m span and 21.46~m$^2$ for 12~m, about 6\% above the values for the adopted positions. The largest values lie at a corner of this grid, with the bus and the radiators at their highest, so placements outside it could give somewhat more. These maxima fall short of the 34.2~m$^2$ needed in $V$ by factors of 1.65 and 1.59, and of the 26.9~m$^2$ needed in $r$ by factors of 1.29 and 1.25. Allowing for an error of 9\%, the shortfall is about 1.46 in $V$ and 1.15 in $r$.

With the published ratios read as albedos instead (stripes 1, radiators 0.53, bus, antenna rear and spheres 0.39, panels 0.125, antenna front 0.06), at most 9.62~m$^2$ can leave the spacecraft for the 10~m span and 10.16~m$^2$ for 12~m, again over the 80 placements. These fall short of what is needed by factors of at least 3.3 in $V$ and 2.6 in $r$, or 3.0 and 2.4 allowing for the 9\% error, at any spin axis and for scattering that is roughly symmetric about the Sun direction. The ratios were, however, fitted with Lambertian surfaces, and a fit with another scattering law could give different ones; the 9\% allowance likewise comes from Lambertian models and is untested for other laws. Because the phase curve is built from all eight nights, the energy bound is valid only where glints play no part on any night, in particular on the May--June nights, which supply most of the integral. For idealized materials with white panel backs (C2) the maximum is the full silhouette, at most 27.0 and 31.5~m$^2$ for the two spans. Allowing for the 9\% error, the energy bound then does not exclude C2 in $r$ for either span, nor in $V$ for the 12~m span; for 10~m it leaves a shortfall of about 1.16 in $V$.

The projected area gives a third bound, which needs no reflectance values. For Lambertian surfaces with reflectance up to 1, direct light cannot exceed the area the spacecraft presents to the Earth or to the Sun, whichever is smaller, with hidden surfaces excluded. At the authors' axis this area is only 17--21~m$^2$ on the low-phase nights (both spans, the 80 placements), against 44--48~m$^2$ observed in $V$. Even with every surface perfectly white, solar cells included, and with the reflections between components added, the model is at least 1.8 times too faint in $V$ (1.7 without November~7) and 1.3 in the conservative variant; for the adopted positions these values are 2.0, 1.9 and 1.4. The reflections are largest when the underside of the bus just clears the torus, a placement that the grid includes.

\subsection{Other checks}
\label{sec:other}
\paragraph{Glint cone.} For case A in the conservative variant and the 10~m span, the minimum over spin axes is 2.42, 2.32, 2.08, 2.07 and 2.05 for cuts of 10, 15, 20, 25 and 30$^\circ$ (12~m span: 2.37, 2.19, 2.05, 2.04 and 2.03). For every cut except 15$^\circ$ the binding night is May~31 or June~5. In $V$ the same cuts give 2.61--3.07 (12~m span: 2.58--3.01).

\paragraph{Phase information.} The bounds use only rotation averages; the light-curve shapes enter only the glint and flash arguments, the reproduction of their fit (Sect.~\ref{sec:fidelity}) and the test of the sphere brightness (Sect.~\ref{sec:results}).

\paragraph{Calibration.} As in Sect.~2, I assume the absolute calibration of \citet{SantanaRos2025}. The one independent calibrated measurement from 2024 that I know of suggests that, if anything, the Source Data are too faint: Keck photometry on June~8 at 93$^\circ$ phase, $R_C=20.72$ \citep{Bolin2025}, converted with the colors of Sect.~2, is 0.4--0.5~mag brighter than the Source Data interpolated to that phase angle between June~5 and 12 (0.7--0.8~mag if the file is in $r'$). If the June nights are too faint by this amount, no shortfall here decreases, but the fading between November~1 and June~12 falls from 13.8 to about 7--9 times, still about three to four times that of the bound at the authors' axis. Between May~21 and June~12 the fading likewise falls from 8.4 to about 4--6 times, still about two to three times that of the bound. I have not assigned an uncertainty to the nightly zero points. Because each shortfall is taken on the worst night, errors in the zero points also bias it upward. For independent errors of 0.1~mag per night, a Monte Carlo simulation gives shortfalls that are on average 2--7\% too high, and 16--20\% too high in one trial in twenty; an error of 0.1~mag is close to the rms scatter of the nightly means around the phase curve (0.10~mag), which also includes changes of aspect. This leaves the conclusion at the authors' axis unchanged, but the minimum over spin axes is better read as about 2 than as at least 2. The mean magnitudes of the light curves scatter by $-0.16$/$+0.23$~mag about the phase curve (Sect.~2), so shortfalls within a few tens of percent of 1 should not be read as firm exclusions.

\section{Discussion}
The published solution matches the shapes but falls short of the brightness by a factor of at least 6.3 in $V$ on the nights they fitted (2.0~mag; 6.7 with November~7), or 4.8 (1.7~mag) in the conservative variant. The shortfall is largest near opposition, where real surfaces can be brighter than Lambertian ones, but it does not rest on those nights: without the three fall nights it is still at least 4.9 in $V$ and 3.8 in the conservative variant, and with only the nights at 61--101$^\circ$ phase at least 2.9 and 2.3 (Sect.~\ref{sec:results}). The approximate energy bound does not assume the Lambertian law, although the ratios and its 9\% allowance come from Lambertian models. With the published ratios read as albedos, the spacecraft reflects about 2.4 times too little light in $r$ at any spin axis, for scattering roughly symmetric about the Sun direction (Sect.~\ref{sec:energy}). At the authors' axis, none of my estimates for mirror-like reflection closes the gap: mirror-like spheres leave a shortfall of at least 3.5, and a generous estimate for every surface about 2. Moving the spin axis to where a glint from the panels could help is not enough for the published brightness ratios: with Lambertian surfaces and cover-glass panels the May--June nights then leave a shortfall of about 2 in the conservative variant, for any cut up to 30$^\circ$ (1.36 without the stripe enhancement, for the 15$^\circ$ cut).

With the published ratios and Lambertian surfaces, only a calibration error could change the conclusion: the 2024 photometry would have to be too bright by more than a factor of 4.8 (1.7~mag) at the authors' axis, or by a factor of about 2 (0.75~mag) at the most favorable other axis. A larger reflecting area would make a different model, which would need its own fit to the light curves. The brightness ratio between the GTC nights of November~1 and June~12 also disagrees with the published model at the authors' axis: the object fades 13.8 times, while the bound and their model as fitted, both Lambertian, fade about 2.2 and 1.1--1.33 times. This ratio depends only on how those two nights are calibrated relative to each other (the zero points of two star fields, and losses from trailing at different rates of motion); reconciling the bound with the object would take an error of about 2~mag between the two nights, against the 0.4--0.8~mag that the Keck comparison of Sect.~\ref{sec:other} suggests for June.

That leaves models other than the published one. Strongly mirror-like surfaces could supply most of the light as a steady glint: the bus, for spin axes near the ecliptic poles, or the panels together with the spheres, for spin axes near the low-phase glint direction. Near-perfect white surfaces everywhere except the solar cells are not excluded either. If they are Lambertian, the bound excludes them near the published axis and its opposite (within 55$^\circ$ for the adopted positions, though only narrowly near that limit). Even if the solar cells are perfectly white too, a Lambertian spacecraft at the published axis falls short by about 1.3 in the conservative variant (Sect.~\ref{sec:energy}), too small a margin to count as a firm exclusion. If these white surfaces are not Lambertian and the panel backs are white as well, the energy bound does not exclude them in $r$, even at the published axis. None of these has been fitted to the light curves. White spheres, like the ADU tanks in their photograph, would leave a shortfall of only about 1.2 in the conservative variant at the best other spin axis, where they have not been fitted either; at the authors' axis the light curves reject them, and the shortfall there is still about 2.1 (Sect.~\ref{sec:results}).

The test uses only the model of \citet{Hibberd2026b} and the 2024 data they fitted, plus the one night they did not fit. That night sets the nominal $V$ values but is excluded from every conservative value. The test does not address the orbital part of the hypothesis \citep{Hibberd2026a,Hibberd2026b} or the object's nongravitational acceleration. Nor does it exclude an artificial origin in general: another spacecraft, or a different model of Phobos~1, would need its own test.

As published, the Phobos~1 model is inconsistent with the calibrated brightness of 1998~KY$_{26}$. If the object is Phobos~1 or another spacecraft, a revised model would have to match both its light curves and this brightness: a rotation-averaged $pA\,\Phi$ of 44--48~m$^2$ at 5--11$^\circ$ phase on the $V$ scale (Table~\ref{tab:nights}), and a drop in brightness by a factor of about 14 between 5$^\circ$ and 101$^\circ$ (about 7--9 if the June nights are as bright as the Keck photometry suggests). The ancillary code can be adapted to test such a model against the same data. With the Hayabusa2\# rendezvous abandoned, remote measurements like these are the main tests in sight. The 1998 Goldstone radar observations \citep{Ostro1999} could provide such a test, with independent data: a radar-scattering model of the spacecraft would have to match the echo spectra, radar cross section and circular polarization ratio reported there. \citet{SantanaRos2025} reprocessed the spectra and compared them with echoes from their light-curve shape models, fitting only the size; the same comparison could be made for a model of Phobos~1. I welcome further tests, photometric or otherwise.

\section*{Acknowledgments}
I thank T.~Santana-Ros and coauthors for publishing their photometry with full geometry, without which this test would not have been possible. I thank A.~Hibberd, A.~Crowl and A.~Loeb for describing their model in enough detail to allow an independent test.

AI disclosure: I prepared the analysis code and an initial draft of the text with the assistance of Claude (Anthropic). I checked the code, the results and the references myself, and I take full responsibility for the content of this paper.

\section*{Data and code}
The analysis scripts and their outputs are included as ancillary files with this arXiv submission:
\begin{itemize}
\item \texttt{ky26\_note.py}, \texttt{ky26\_ms.py} and \texttt{ky26\_c2.py} hold the model, the reflection calculation and the cases;
\item \texttt{ky26\_robust.py} gives the values at the authors' spin axis and the energy bound for the adopted positions;
\item \texttt{ky26\_fine.py}, \texttt{ky26\_hybrid.py} and \texttt{hires\_edge.py} carry out the search over spin axes and its checks;
\item \texttt{ky26\_barrel.py} and \texttt{ky26\_reach.py} treat glints from the bus and find where the bound stops excluding each case;
\item \texttt{ky26\_specular.py} and \texttt{ky26\_spheres\_spec.py} bound mirror-like reflection at the authors' axis and add mirror-like spheres to the search;
\item \texttt{table2.py} assembles Table~\ref{tab:results}, and \texttt{text\_numbers.py} most of the other numbers quoted in the text;
\item \texttt{figures.py} produces the figures;
\item \texttt{test\_note.py} contains the unit tests;
\item \texttt{checks/} holds the reproduction of the published light-curve fit (Sect.~\ref{sec:fidelity}), the search over the 68\% region of the authors' pole, the values at the authors' axis at higher resolution, the energy and projected-area bounds with their scan over placements, the values of Table~\ref{tab:results} for other placements, the test of the energy-bound assumption (Sect.~\ref{sec:energy}), the estimates for mirror-like surfaces, the variants with white spheres and the light-curve test of their brightness (Sect.~\ref{sec:results}), the path-traced reflections (Sect.~\ref{sec:multi}), the case with the bus bottom lit (Sect.~3) and the Monte Carlo of the nightly zero points (Sect.~\ref{sec:other});
\item a \texttt{README}, the JSON result files and the reflection grids are included.
\end{itemize}
The photometry is the Source Data of \citet{SantanaRos2025}, file\\ \texttt{41467\_2025\_63697\_MOESM3\_ESM.txt} (doi:10.1038/s41467-025-63697-4).

\end{document}